\documentclass[preprint2]{aastex}

\slugcomment{To appear in PASP}

\shorttitle{Slitless spectroscopy with HST/ACS} 
\shortauthors{Pasquali et al.}

\begin{document}


\title{Slitless grism spectroscopy with the HST \altaffilmark{1} Advanced 
       Camera for Surveys}


\author{A. Pasquali}
\affil{Institute of Astronomy, ETH H\"onggerberg, HPF, CH-8093 Z\"urich,
       Switzerland \\ 
       Max-Planck-Institut f\"ur Astronomie, K\"onigstuhl 17, D-69120 
       Heidelberg, Germany}
\author{N. Pirzkal}
\affil{Space Telescope Science Institute, 3700 San Martin Drive,
       Baltimore, MD 21218, USA }
\and
\author{S. Larsen, J.R. Walsh, M. K\"ummel}
\affil{ESO/ST-ECF, Karl-Schwarzschild-Strasse 2, D-85748 Garching
       bei M\"unchen, Germany}


\altaffiltext{1}{The Hubble Space Telescope is a project of international
cooperation between NASA and the European Space Agency (ESA).
The Space Telescope Science Institute is operated by the Association
of Universities for Research in Astronomy, Inc., under NASA Contract
NAS5-26555.}


\begin{abstract}
The Advanced Camera for Surveys on-board {\it HST} is equipped with a 
set of one grism and three prisms for low-resolution, slitless spectroscopy  
in the range 1150 \AA\/ to 10500 \AA. The G800L grism 
provides optical spectroscopy between 5500 \AA\/ and 
$>$ 1 $\mu$m with a mean dispersion of 39 \AA/pix and 24 \AA/pix 
(in the first spectral order) when coupled
with the Wide Field and the High Resolution Channels, respectively.
Given the lack of any on-board calibration lamps for wavelength and
narrow band flat-fielding, the G800L grism can only be calibrated
using astronomical targets. In this paper, we
describe the strategy used to calibrate the grism in orbit, 
with special attention to the treatment of the field dependence of 
the grism flat-field, wavelength solution and sensitivity in both
Channels.
\end{abstract}




\section{Introduction}
Studies of galaxy formation and evolution at high redshift greatly
benefit from multi-object spectroscopy, which is widely accessible from
the ground. How far in lookback time one can go is, though, severely
limited by the emission spectrum of the night sky. This can be overcome
with observations from the space, by performing multi-object spectroscopy
which turns out to be rather sensitive to compact sources. Indeed, as shown by 
the GRAPES program (Pirzkal et al. 2004),  grism spectroscopy with the HST
Advanced Camera for Surveys (ACS) can detect the continuum emission
of sources as faint as z$_{AB} \sim$ 27 in 40 orbits. Multi-object 
spectroscopy with HST comes with low spectral resolution
to compensate for the telescope size, and without slits. As we will discuss
in the next Sections, the absence of a slit allows spectra of different
sources to overlap, the size of the source to decrease the instrumental
spectral resolution and the background to increase since each pixel collects
the background emission integrated over the whole spectral range of the
dispersing element.

Slitless spectroscopy has been made available with the Hubble Space
Telescope (HST) since its launch, in 1990. The first instruments to
be equipped with grism/prism elements were the Wide Field Planetary
Camera 1 (WFPC1) and the ESA Faint Object Camera (FOC).  
WFPC1 contained an UV grism (G200L, Horne \& MacKenty 1991), covering 
the spectral range between 1300 \AA\ and 4000 \AA\ with a resolving power 
of about 100 (at $\lambda$ = 3000 \AA) in the first order. The grism was not extensively 
calibrated in wavelength and flux and was never used for GO science
observations. During the time of WFPC1, HST was not yet corrected for
spherical aberration.
\par
Two prisms were mounted in the FOC; they covered
the spectral ranges 1150 \AA\ $\leq \lambda \leq$ 6000 \AA\ 
and 1600 \AA\ $\leq \lambda \leq$ 6000 \AA\
with a resolving power R $\simeq$ 100 at $\lambda$ = 1500 \AA\ and 
$\simeq$ 250 at $\lambda$ = 2500 \AA, respectively. They provided 
slitless, FUV and NUV spectroscopy across the F/96 (14$'' \times$ 14$''$)
and F/48 (28$'' \times$ 28$''$) fields of view. 
Both prisms were supported modes of FOC, i.e. they were
calibrated in wavelength and flux before launch and in-orbit
(cf. Paresce \& Greenfield 1989, Hack 1992, Hack 1996). 
Nevertheless, the FOC prisms were seldomly requested by 
users during the 7 year life-time of this instrument. The
FOC prisms were used to
address a number of scientific cases, such as the measurement of
the UV flux shortward of $\lambda$ = 1800 \AA\ for 
a sample of QSOs at $z >$ 3
(Jakobsen et al. 1993); the study of the UV morphology and expansion
of the SN 1987A ejecta (Jansen \& Jakobsen 2001); and to spatially 
resolve the spectra of the giant stars in the Capella binary
system (Young \& Dupree 2002).   
\par
Since 1997 slitless spectroscopy has become feasible also at infrared
wavelengths with NICMOS Camera 3, which is equipped with three grisms
(G096, G141 and G206) covering the spectral range 0.8 - 2.5 $\mu$m at a
resolving power of $\simeq$ 200 (Noll et al. 2004).  
While the first two grisms exploit the
low background of HST, G206 is subject to the thermal background
emission from HST. The three grisms provide multi-object spectroscopy
over a field of view of 51$'' \times$ 51$''$ at a spatial resolution of
0.2$''$/pix, and were mainly used to study the evolution of field galaxies at
0.7 $< z <$ 1.9 (McCarthy et al. 1999) and to investigate the nature of planets 
and Kuiper belt objects (McCarthy et al. 1998).
During the first two years of NICMOS, the grisms were calibrated in
wavelength and flux in orbit, by observing a Galactic Planetary Nebula
and a White Dwarf\footnote{cf. http://www.stecf.org/newsletter/webnews1/nicmos}.
Software for extracting spectra (NICMOSlook) was released by
Freudling \& Pirzkal (1997) and Freudling (1999).  
\par\noindent
The NICMOS grisms have been quite extensively used, by both GO proposals
and parallel programs. Prior to the Servicing Mission 3B, the
accepted GO proposals requiring spectroscopy represented about
14$\%$ of all the successful NICMOS proposals, while in the case of FOC the
approved proposals making use of the prisms amount to only $\simeq$ 4$\%$ of
the total.

Although designed as a slit, high-to-medium resolution spectrograph,
STIS has also provided slitless spectroscopy mostly with the G750L
grating (R $\simeq$ 750 at 7500 \AA) when employed in parallel mode. An 
IRAF routine (SLITLESS in STSDAS/STECF) was
developed by N. Pirzkal for the extraction of slitless spectra.
\par\noindent
The slitless parallel data of STIS have been analyzed to 
search for new low-mass stars and brown dwarfs (Plait 1998a and 1998b),
to study the intermediate-redshift Universe (de Mello \& Pasquali 2001,
Teplitz et al. 2001, Teplitz et al. 2003) and to derive the hydrogen 
reionization edge (Windhorst et al. 2000). 

In March 2002, HST was refurnished with a new instrument,
the Advanced Camera for Surveys (ACS) which offers the unique
combination of a large field of view (3$'$.4 $\times$ 3$'$.4) with a 
high angular resolution (0.05$''$/pix, cf. Sect. 2). The main task
of ACS is to perform deep imaging and spectroscopic surveys.
In contrast to older generation
instruments, the ACS grism and prisms have been extensively calibrated
in laboratory and specific GO calibration proposals have been set up
to assess their in-orbit performance. It is believed that such an
effort will involve a larger number of HST users in using the ACS
spectroscopic capabilities and will return relevant scientific results.
In the first two years of ACS, the grism has been extensively used
for the search for supernovae type Ia at high redshift (Magee et al. 2002, 
Tsvetanov et al. 2002, Blakeslee et al. 2003,
Riess et al. 2004) and for high-redshift Ly$\alpha$ emitters, and to 
study galaxy formation
and evolution at intermediate-to-high redshifts, especially in the
UDF field (Pirzkal et al. 2004,2005a,b; Xu et al. 2005; Daddi et al. 2005;
Malhotra et al. 2005; Pasquali et al. 2005a,b; Rhoads et al. 
2005).

In this paper, we review the concepts and technique of the grism calibrations, 
which have been developed for non-drizzled, geometrically distorted data. 
Drizzling\footnote{Drizzling is a new technique for the linear combination of 
images known formally as variable-pixel linear reconstruction. It is based 
on a continuous set of linear functions that vary smoothly from the optimum linear 
combination technique -- interlacing -- to the old-standby, shift-and-add
(cf. Fruchter \& Hook  1998, Hook \& Fruchter 2000).}  
the grism data, {\it before establishing their dispersion
solution}, involve a
change of pixel size and geometry which in turn affect the
dispersion and zero point of the wavelength calibration and also the amplitude
and wavelength dependence of the flat-field correction for grism spectra. 
These quantities depend on the drizzle parameters which can be specified 
differently each time by different users. For these reasons, the grism
calibration has been performed only for raw grism data.

\section{The ACS instrument layout}
The ACS is a three-channel instrument which performs broad and
narrow band filter imaging from UV to NIR, polarimetric imaging 
at ultraviolet and optical wavelengths, and low-resolution slitless
spectroscopy in the range from $\simeq$ 1150 \AA\ to $\lambda$  $\simeq$ 
1 $\mu$m with a set of grism and prisms (cf.
http://www.stsci.edu/hst/acs/documents). The ACS spectral
elements can be summarized as follow: 
\par\noindent
The Wide Field Channel (WFC) has a 3$'$.4 $\times$ 3$'$.4 
field of view at a spatial resolution of 0.05$''$/pix and is
equipped with a grism (G800L) covering the spectral range from 5500 \AA\ 
to 1 $\mu$m
with a response peak at $\simeq$ 7200 \AA\ and $\simeq$ 6000 \AA\  in the 
1$^{st}$ and 2$^{nd}$ orders,
respectively. The grism resolving
power  is  $\simeq$ 100 and the dispersion is nearly linear,
$\simeq$ 39 \AA/pix in the 
1$^{st}$ order and $\simeq$ 20 \AA/pix in the 2$^{nd}$ order. The tilt of 
the spectra is nearly -2$^o$ from the image X axis, varying slightly across
the field.
\par\noindent
The High Resolution Channel (HRC) is characterized by 
a 26$'' \times$
29$''$ FOV with a spatial resolution of 0.027$''$/pix and uses
the same grism of the WFC, which introduces a tilt in the spectra of -38$^o$ 
(from the image X axis) and offers a dispersion of $\simeq$ 24 \AA/pix 
in the 1$^{st}$ order and of $\simeq$ 13 \AA/pix in the 2$^{nd}$ order. The HRC 
also features a prism (PR200L) covering
the 1600 - 5000 \AA\ range, with a non linear dispersion varying from 2.6
\AA/pix at 1600 \AA\ to 105 \AA/pix at 3500 \AA\ and 560 \AA/pix at 
5000 \AA.
\par\noindent
The Solar Blind Channel (SBC) has a field of view of 31$'' \times$
35$''$ where each pixel is 0.032$''$ in size. Two prisms are
available with the SBC: PR110L covering from 1150 \AA\ to 2000 \AA, and
PR130L from 1220 \AA\ to 2000 \AA (i.e. geocoronal Ly$\alpha$ is blocked).
The dispersion of PR110L is  2.6 \AA/pix at 1216 \AA,
while the dispersion of PR130L varies from 1.65 \AA/pix at 1250 \AA\
to 20.2 \AA/pix at 1800 \AA.

The ACS grism and prisms are used in slitless mode. This raises 
a number of important issues:
\par\noindent
{\it i)} The extension of a source along the dispersion axis
is given by the spatial profile of the object, and affects the
achievable spectral resolution.
\par\noindent
{\it ii)} The zero point of the grism wavelength calibration is defined 
by the position of the object in the direct image. When no direct image
of the source is available, the position of the 0$^{th}$ order can
also be used, though in a less accurate way since the 0$^{th}$ order is
itself dispersed.
\par\noindent
{\it iii)} The spectroscopic flat-field is field and wavelength
dependent. Depending on the position of a target in the direct
image, the same pixel in the grism/prism image receives different wavelengths.
\par\noindent
{\it iv)} Each pixel in a grism/prism spectrum has a contribution from
the background emission which is integrated over the entire spectral
range of the grism/prism.
\par\noindent
{\it v)} Depending on the luminosity of a source, the grism orders
higher than 1 (at higher X coordinates in the grism image) and the grism 
negative orders (at lower X coordinates) can be identified. No
filter can be used to suppress higher and negative orders. 
\par\noindent
{\it vi)} Depending on the position of the object in the direct image,
the source spectrum can be truncated in the grism/prism image. Objects outside
the field of view can still produce spectra in the grism/prism image.
\par\noindent
{\it vii)} The extension of grism spectra and the number of their
orders can give overlap and contamination among spectra of adjacent
objects.
\par\noindent
Therefore, the wavelength calibration of the ACS grism and prisms requires
a full description of the grism/prisms physical properties,
such as the tilt
of the spectra, the length of the prism spectra and of 
the grism orders together with their separation
from the grism 0$^{th}$ order, the (X,Y) offsets between the object
position in the direct image and the corresponding 0$^{th}$ order in the
grism image. Indeed, these quantities allow the identification of the
spectrum in the grism/prism image of any object in the direct image, the
tracing of spectra and set the size of the aperture extraction for each
prism spectrum and grism order (Pasquali et al. 2003a,b).
As described in the next Sections, these parameters are 
field-dependent because of the geometric distortions of ACS. This is
also true for the flat-field, which an accurate flux calibration of 
the spectra depends on.

\section{The effective spectral dispersion of the ACS grism}
Since no slit is available, the nominal dispersion of the ACS grism 
gets convolved with the object size and, in the
case of non-round sources, with the object size along the dispersion axis as
determined by the object orientation. This convolution obviously
lowers the grism resolution and plays a key role in planning
the spectroscopic calibrations and the GO observations 
with the ACS (Pasquali et al. 2001a). 

To quantify the effect of object size and orientation on the
effective dispersion of the grism, we have simulated the optical spectrum
of the Galactic Planetary Nebula NGC 7009 using SLIM 1.0 (Pirzkal
\& Pasquali 2001, Pirzkal et al. 2001b\footnotemark \/).
\footnotetext{http://www.stecf.org/instruments/acs/slim} 
SLIM 1.0 is a slitless spectroscopy simulator developed by
N. Pirzkal and is designed to produce realistic and photometrically
correct spectra acquired with the ACS grism and prisms. Briefly, any
input spectrum is redshifted as required and scaled to a desired magnitude 
in a reference filter, then dispersed according to the grism/prisms
dispersion solution. At each wavelength, the dispersed spectrum is
convolved with the grism/prisms throughput curve and with a gaussian PSF whose
FWHM value is set by the HST diffraction limit at any specific wavelength.

Two sets of simulations have been produced for both the WFC and the HRC 
which do not include background or read-out noise. The first set is based on
the assumption that NGC 7009 is disc-shaped, with uniform surface brightness 
and diameter varying 
from 0$''$.05 to 0$''$.1, 0$''$.4 and 1$''$ for the WFC, and 
from 0$''$.06 to 0$''$.2 and 0$''$.5 for the HRC. 
The second set assumes that NGC 7009 is elliptical with a
varying position angle (PA) with respect to the
dispersion axis. The nebula size is 0$''$.1 $\times$ 0$''$.25 and
0$''$.05 $\times$ 0$''$.12 for the WFC and the HRC respectively. The
PA of the nebula major axis is assumed to vary from 0$^o$ to 20$^o$,
45$^o$ and 90$^o$ so that
the object major axis progressively goes from being aligned along to
being perpendicular to the dispersion axis. 
\par\noindent

We have measured the FWHM of a few
lines in the 1$^{st}$ and 2$^{nd}$ order spectra obtained
for the WFC and the HRC in both sets of simulations, in order to quantify 
the {\it degradation}
of the grism nominal dispersion as a function of object size. 
These FWHM values have been normalized to the FWHM of a source with
a 0$''$.05 diameter and are plotted in Figure 1 as a function of
the object size along the dispersion axis in arcsecs, for
both the WFC and the HRC. As expected, 
the line FWHM  nearly doubles by doubling the
object linear size, implying that the 
grism dispersion degrades by almost the same factor as the
object size increases. The flattening in the curves for linear sizes
larger than 3 $\times$ 0$''$.05 for the WFC (3 $\times$ 0$''$.03 for
the HRC) is due to severe line blending which
makes the line width measurements unreliable.

\section{The calibration strategy}

\subsection{Ground calibrations}
During the ground testing of ACS at the Goddard Space Flight Center,
the Refractive Aberration Simulator (RAS/RAMP) was used to acquire a
series of grism spectra for both the WFC and the HRC. Specifically,
spectra were obtained through a 4 $\mu$m pinhole aperture fed by an 
Argon lamp and QTH fiber lamp sources. Spectra were acquired at
different positions across the field of view, 
located close to the center and to the CCD amplifiers in the corners. 
These data provided a first wavelength calibration,
and also allowed to check the amplitude of the field
dependence of the grism resolution. The spectra of the continuum lamp
were used to derive the relative, integrated throughput of the different grism
orders across the field of view, but were not suited to determine
the throughput as a function of wavelength for the various grism orders.
We could then establish that the total power in the 0$^{th}$ order is
2.5$\%$ of that in the 1$^{st}$ order, and 1$\%$ and 0.5$\%$ in the positive 
and negative higher orders, respectively.
\par\noindent
Since the grism is sensitive to wavelengths redder than 5500 \AA, we
could identify and use only the Argon lines redward of 6700 \AA.
This made the wavelength solutions of the grism orders less accurate than 
those derived in orbit (cf. Sects. 9 and 10). For example, the wavelength
solution of the positive 1$^{st}$ order could only be fitted with a first-order
polynomial, which, later on, gave raise to significant wavelength offsets in the
spectra of the in-orbit calibrators.

\subsection{In-orbit calibrations}
The results obtained in Sect. 3 pose important constraints on the planning 
of wavelength
calibration observations,  such as those foreseen for the Servicing Mission 
Orbital Verification (SMOV, following the installation of ACS onboard
HST) tests and
subsequent GO calibration programs. ACS is not equipped with arc lamps, 
therefore the in-orbit dispersion correction of the grism can only be
determined from observations of specific astrophysical sources. 
For this purpose, calibration targets should satisfy the following
requirements:
\par\noindent
{\it i)} the spatial extension of the target should be minimal,
to allow accurate determination of the dispersion and wavelength zero point; 
\par\noindent
{\it ii)} the target brightness allows for reasonably short exposure times;
\par\noindent
{\it iii)} the target spectra should contain a significant number of 
unblended emission lines; 
\par\noindent
{\it iv)} no extended nebulosity is associated with the targets which would
degrade the dispersion of the instrument and increase the local 
background;
\par\noindent
{\it v)} the spectrophotometric variability (either intrinsic or induced by
an eclipsing companion or by orbital motion) is negligible so that 
emission features can be identified at the same wavelength at any 
observational date;
\par\noindent
{\it vi)} the targets should not lie in crowded fields to avoid contamination
by nearby spectra.

Spectroscopy in slitless mode relies on a pair of direct and grism images,
since the zero point of the grism wavelength solution is set by the 
position of the object in the direct image. Because the ACS is tilted with
respect to the optical axis of HST, its field of view suffers severe geometric 
distortions
which introduce a field dependence in the grism dispersion and wavelength 
zero point.
Hence, calibration targets have to be observed at a number of
positions across the WFC and HRC chips in order to be able to parameterize the
field dependence of the grism wavelength solution. This need strengthens 
requirement {\it i)}: the exposure times for the calibration targets should
be reasonably short to allow multiple acquisitions across the field of view. 

In what follows, we discuss the observational strategy adopted for
the SMOV tests and following GO calibrations in order to achieve 
the wavelength calibration of the ACS grism. 

\subsection{The traditional calibrators: Planetary Nebulae}
At optical wavelengths, the spectra of Planetary Nebulae  
are dominated by strong, narrow emission lines mainly due to H, HeI, 
[O~II], [N~II], [O~III] and [S~III]. The [O~III] and [~NII] lines are often used to measure
the expansion velocity of Planetary Nebulae (PNe); Galactic PNe
span a wide range in expansion velocities, from  $\sim$ 5 
km s$^{-1}$ to 50 km s$^{-1}$ (Acker et al. 1992). In the worst case,
an expansion velocity of 50 km s$^{-1}$ corresponds to a FWHM value
of $\sim$ 1 \AA\/ at the restframe H$\alpha$ line. 
With a resolving power of R $\simeq$ 100, the ACS grism does not resolve any
Planetary Nebula in velocity and any Planetary Nebula would be equivalent 
to a ground arc lamp. Among all the known PNe, only those
which are compact (with a full size smaller than 0.1$''$) can be observed
with the ACS grism according to the requirements listed in Sect. 4. PNe 
in the Galaxy and in the Magellanic Clouds
are rarely compact when imaged with HST and even when compact often
have a low intensity halo. Therefore, in order to preserve the nominal 
dispersion of the ACS grism, we have to resort to more distant PNe,
for example those in M31.

To assess the feasibility of grism observations of extragalactic PNe
in terms of exposure time and S/N ratio in the lines, we simulated
the ACS spectra of the brightest and most compact PNe in M31 using
SLIM 1.0 (cf. Pasquali et al. 2001b). The selected PNe  were originally
studied by Ciardullo et al. (1989) and Jacoby \& Ciardullo (1999).   
Our simulations show that PNe can be used to determine the dispersion
correction of the grism 1$^{st}$ order, but fail the
wavelength calibration of the 2$^{nd}$ since only two lines can be 
detected with a reasonable S/N ratio.
To improve this, the integration time should be longer by a factor of $\sim$3
(equivalent to $\sim$ 1 h per pointing),
thus making PNe too time-consuming for routine GO calibration programs.
An additional disadvantage of observing PNe in M31 is the field
crowding, which makes spectra of adjacent sources overlap.
Nevertheless, PNe can be employed as secondary targets, to verify
the accurateness of the wavelength calibration.

\subsection{An alternative choice: Galactic emission line stars}
Having discharged PNe, our simulations show that  
Galactic Wolf-Rayet (WR) stars satisfy the calibration requirements. 
As Figure 2 points out, WR stars
of spectral type WC6 - WC9 have a large number of bright emission
lines (He and C) in the spectral range covered by the ACS grism. 
The only drawback here is that the velocity of their stellar winds 
considerably broadens their emission features. 

The known Galactic WR stars (van der Hucht 2001) have wind
velocities between $\sim$ 700 km s$^{-1}$ and 3300 km s$^{-1}$ with a
mean wind speed of about 1700 km s$^{-1}$ and with 19$\%$ of the sample having
a wind velocity larger than 2100 km s$^{-1}$. We thus assumed a wind 
speed of 2000 km s$^{-1}$ and computed the line broadening induced in  
WFC and HRC grism spectra of WR stars. 
The results indicate that such a stellar wind 
is hardly resolved in the 1$^{st}$ order spectra and 
partially affects the 2$^{nd}$ order spectra. Higher wind velocities
rapidly decrease the grism resolution at any order.

Clearly, the wind velocity constraint ($v_{\infty} \le$ 2000 km 
s$^{-1}$) has to be added to the selection criteria discussed earlier in
Sect. 4, together with the requirement that the variability of the
selected targets is negligible. Spectro-photometric variability could
change the number of detectable emission lines, and variability in the
stellar wind could change the line FWHM; both cases would make the
grism wavelength calibration less accurate.
From the catalogue of van der Hucht (2001),
we have selected two stars, WR45 and WR96, whose 
spectral type, coordinates, V magnitude
and wind velocity are listed in Table 1. We produced SLIM 1.0 spectra based
on two WR templates kindly provided by P. Crowther and closely matching
the spectral types of WR45 and WR96. The simulations
are plotted in Figures 2 and 3 for the WFC and the HRC 1$^{st}$ and
2$^{nd}$ orders, respectively. In the case of the WFC simulations, an
exposure time of 10 s and 60 s has been assumed for the 1$^{st}$ and
2$^{nd}$ order, respectively, while for the HRC 20 s and 60 s. 
Contrary to the Planetary Nebula case,
the simulated spectra of WR45 and WR96 indicate that a relatively large
sample of emission lines can be used for the 
wavelength calibration of both the 1$^{st}$ and 2$^{nd}$ orders of the ACS
grism.

As a counter-check, the simulated spectra of WR45 and WR96 have been
used to derive the grism wavelength solution for both the WFC and the HRC.
We have measured the position of each identified line in the SLIM output 
in pixels scaled to the position of the object in the direct image, and fitted
pixels against wavelength assuming a first order
polynomial. The accuracy of the dispersion correction thus obtained
is better than 0.3 pixels for both the 1$^{st}$ and 2$^{nd}$ order
(i.e. 12 \AA\/ and 7 \AA, respectively). 

\subsection{ESO/NTT support observations of WR45 and WR96}
Since no high resolution spectra were available in literature
at the time of the ACS ground calibrations, we observed WR45 and
WR96 with the EMMI spectrograph, mounted on the ESO NTT telescope
(266.D-5653, PI Pasquali). The observations were performed in service mode
in June and August 2001 as part of the ESO Director General
Discretionary Time. Spectra were acquired through the REMD grating
$\#$ 8 (between 5000 and 7500 \AA\/ at 1.26 \AA/pix, and in the
range 7300 - 9750 \AA\/ at 1.26 \AA/pix) and the RILD grism $\#$ 2
(from 4000 to 9000 \AA\/ at 2.7 \AA/pix). The high resolution
spectra were used for an accurate identification of the He and C
lines typical of WR stars, while the low resolution spectra provided
the flux calibration of the stellar continuum and lines.
The NTT, high resolution spectra of WR45 and WR96 are plotted in
Figure 4. They were used as inputs to SLIM 1.0 
in order to simulate the WFC and HRC grism observations of WR45 and
WR96 and to optimize the exposure time and S/N value of the
in-orbit calibrations.

\section{The ACS observations of WR45 and WR96}
As part of the SMOV and INTERIM calibration programs (9029 and 9568,
PI Pasquali), we observed
WR45 at a number of positions across the
field of view of the WFC and the HRC. WR96 was also observed, but only
with the WFC, at the same positions used for WR45.
We selected 10 pointings for the WFC 
(among which the five from the ground tests) and
5 (only those observed during the ground calibrations) 
for the HRC, in order to sample the field dependence of the grism
properties, which ground-based calibrations indicated to be 
smaller for the HRC.
WR96 was re-observed in Cycle 12 together with the Planetary Nebula
SMP-81 in the LMC
(Program 10058, PI Walsh). These observations were designed to provide
additional sampling points across the field of view to better map the spatial 
variation of the wavelength solution. The INTERIM, and Cycle 12 data
(i.e. the object positions in the direct image) are illustrated in 
Figures 5 and 6.

At each
pointing a pair of direct and grism images was acquired and repeated
at least twice, either during the same visit or in different visits, to
test the repeatability of the filter wheel positioning. Direct images
were taken with the F625W and F775W filters to test whether the target
position in the direct image is filter-dependent. Exposure times 
for imaging and spectroscopy are listed in Table 2 for both Channels and
targets. In particular, the spectroscopic exposure 
times were tuned to get a S/N ratio $\geq$ 30 in both the positive and negative
1$^{st}$ and 2$^{nd}$ orders, and turned out to be long enough to detect
the 3$^{rd}$ and -3$^{rd}$ orders. 

The observations included also two White Dwarfs, GD 153 and G191B2B,
which were acquired at the same positions across the chips as the 
WR stars. Their spectra were used to derive the grism throughput and
flux calibration of the WFC and the HRC grism configurations.

\section{The WFC/grism configuration}
The spectral tilt can be derived from the (X,Y) coordinates of
the line-emission peaks, as measured along the spectrum from the
-3$^{rd}$ to the +3$^{rd}$ order in the grism spectra. We have
fitted up to 20 (X,Y) pairs with a first order polynomial, whose
slope is the tilt of the grism spectra with respect the X axis
of the grism image. 

On average, the spectral tilt in the WFC/grism configuration is
1$^o$.98 $\pm$ 0$^o$.34 from the image X axis, but it varies by 
1$^o$ across the field of view of the WFC (cf. Figure 8). 
The maximum variation ($\sim$ 1$^o$.1) occurs along
the diagonal from the top left corner to the bottom right; 
a variation of 0$^o$.79 is seen along the diagonal from the bottom
left to the top right corner of the field of view (cf. Pasquali
et al. 2003a).

The spectra appear off-set from the object in the direct image. 
We measured the positions of the 0$^{th}$ order 
and the object in the direct image using their
light centroid. We used
the fits of the spectral tilt to calculate the Y$'$ coordinate
on the spectrum corresponding to the X coordinate of the object
in the direct image. The difference (Y$_{0^{th}}$ - Y$'$) represents
an offset between the object in the direct image and the grism
0$^{th}$ order, which allows to locate the spectrum of a source
in the grism image once the source coordinates have been measured
in the direct image.

The mean value of this Y-offset across the field of view is -3 pixels,
although the Y-offset is itself field dependent.
The offset of the object in the direct image from the grism
0$^{th}$ order along the X axis is almost constant across
the WFC field of view and is of about 112 pixels (cf. Pasquali et
al. 2003a and Figure 8).

To estimate the length of the grism orders along the image X axis,
a threshold has
to be set in the counts which distinguishes between pixels of
background and pixels of source spectrum. We set this threshold to
3$\sigma$ level above the background, and measured the X
coordinates of the blue and red edges of each order spectrum.
These points give an estimate of the length and separation of the grism
orders, which show negligible field dependence. 
The 0$^{th}$ order turns out to be dispersed over 23 pixels and
its FWHM is about 4 pixels. The 1$^{st}$ order is about 156 pixels
long, while the 2$^{nd}$ is $\sim$ 125 pixels in length. It should
be about two times more extended than the 1$^{st}$ order, but the
very low throughput of the grism in the 2$^{nd}$ order prevents 
its detection at $\lambda >$ 8000 \AA\/ (cf. Figure 8). 
The -1$^{st}$ and -2$^{nd}$ orders are 102
and 111 pixels, respectively. Note that there is contamination in 
the first order spectrum at $\lambda >$ 10000 \AA\/ by the grism
second power.

During the analysis of the combined INTERIM and Cycle 12 data, we
used a different method to trace the spectral orders. The centroid was
measured directly on the images in the cross-dispersion direction 
along the spectral traces, and the run of Y-offset versus X position 
with respect to the target position in the direct image
was fitted with a straight line. We found this method to provide
nearly identical results to the approach described above.

\section{The HRC/grism configuration}
The mean spectral tilt across the HRC field of view is -38$^o$.19
$\pm$ 0$^o$.12 from the image X axis (cf. Figure 9). 
The spectral tilt decreases
from the top left to the bottom right corner by $\sim$ 0$^o$.05, and 
by $\sim$ 0$^o$.33 along the diagonal from the bottom left to the top
right corner of the field of view (Pasquali et al. 2003b).

The Y-offset is on average -1.5 pixels with little field dependence,
while the X-offset between the grism 0$^{th}$ order and the object
in the direct image is 145 pixels (as shown in Figure 9).

Both the grism 1$^{st}$ and 2$^{nd}$ orders are about 195 pixels 
along the spectral trace, while the -1$^{st}$ and -2$^{nd}$ orders are
$\sim$ 183 and 174 pixels, respectively. As for the WFC/grism configuration,
the sensitivity of the 2$^{nd}$ order redward of 8000 \AA\/ is so low 
that the 2$^{nd}$ order spectrum is not detected any longer 
and turns out to be as extended as the 1$^{st}$.  
The 0$^{th}$ order is dispersed over 9 pixels along the spectral trace 
and its FWHM is 6 pixels.

\section{The method for calibrating in wavelength}
The wavelength calibration of the grism spectra of WR45 and WR96 relies
on the set of template spectra taken for these two stars at medium resolution
with the ESO NTT/EMMI spectrograph. 

Two different approaches were adopted for determining the wavelength solutions.
We first summarize the analysis of the SMOV and INTERIM data, which is
described in more detail in Pasquali et al.\ (2003a). In this case,
the WFC/HRC grism spectra were extracted with the ST-ECF package aXe (Pirzkal et al. 
2001a, Pirzkal et al. 2003a,b),
adopting an extraction aperture 2 pixels wide (to sample the peak of
the instrument PSF) and the background was estimated  
100 pixels away from the spectral trace. Since no wavelength solution had
been defined at this stage, the wavelength scale of the extracted spectra
was in units of pixels along the spectral trace, relative to the
position of the object in the direct image.  Seven emission lines can be 
identified in the extracted $\pm$1$^{st}$ and 3$^{rd}$ orders, while three 
to five lines are identified in the $\pm$2$^{nd}$ and -3$^{rd}$ orders. We
measured for the WFC a mean line FWHM$_{WR}$ of 3, 4 and 6 pixels in
the $\pm$1$^{st}$, $\pm$2$^{nd}$ and $\pm$3$^{rd}$ orders, and for
the HRC 4 and 6 pixels for the $\pm$1$^{st}$ and 2$^{nd}$ orders 
respectively.

The NTT template spectra were convolved with a gaussian function whose $\sigma$
was set to FWHM$_{WR}$/2.36 in order to mimic the ACS grism spectra.
We reidentified the emission line in the new ``degraded'' templates, and
measure their centroid wavelength via multi-gaussian fitting and deblending.
At the same time and with the same procedure,
the line centroids were measured in the WFC/HRC grism spectra in units of
pixels. This method
produced a table of line peak wavelengths and pixels for
each grism order, which were then fitted to derive the wavelength solution
of each spectral order.
Clearly, the order of the fit depends on
the number of identified lines available and on the S/N ratio in the extracted
WFC/HRC grism spectra.  

For the analysis of the INTERIM and Cycle 12 WFC observations of WR96 
(Larsen \& Walsh 2005) the grism spectra were again extracted with aXe as before, except that
an extraction box width of 10 pixels was used. The wavelength solutions
were then determined by fitting the NTT spectra directly to the ACS grism
spectra. The AMOEBA routine (Press et al.\ 1992) was used to minimize
the r.m.s. difference between the smoothed NTT spectra and the grism
spectra as a function of the smoothing length and the wavelength solution
coefficients. Since the ACS spectra were not flux calibrated at this
stage, we also fitted for a wavelength-dependent sensitivity curve,
approximated for this purpose as a 4th order polynomial. 
Figure 7 shows the smoothed NTT spectrum of WR96 and the
ACS grism spectrum for the best-fitting wavelength solution. Clearly,
the fit is not perfect, perhaps partly due to real temporal variations
in the spectra of the WR star. Repeated iterations of the AMOEBA
fits were made, each with slightly different initial guesses for the
fit parameters to test the repeatability of the fits. Generally, the 
wavelength zero-points reproduced within a few \AA\ and the dispersions 
to within 0.1 \AA\ pixel$^{-1}$.
All these procedures were applied to each grism order for each position across 
the field of view.

\section{The in-orbit wavelength solutions of the WFC grism}

An example of 2D grism images taken with the WFC and the HRC is shown
for WR96 in Figures 8 and 9 respectively, where
the different grism orders have been also identified. In these Figures we 
compare the location of the grism orders with the object position in the direct image.
The WR45 spectra for the 
positive and negative orders taken at the center of Chip 1 are
plotted in Figure 10, in units of counts vs wavelength in \AA. 

In the case of the WFC/grism observations, six orders (from the -3$^{rd}$ to
the +3$^{rd}$ orders) were detected together with the 0$^{th}$ order.
Their wavelength solutions computed
for the pointings across the WFC field of view are summarized in
Tables 3 and 4. The blank entries indicate
that the order was not detected because it fell outside the
physical area of the chips. These wavelength solutions have been
computed from the combined INTERIM and Cycle 12 calibration data
(Larsen \& Walsh 2005). They
are reassuringly similar to those obtained from the SMOV/INTERIM data
(that are omitted, for this reason, from this paper. For more
details cf. Pasquali et al. 2003a). Indeed, the difference between the
two calibrations is generally less than one pixel. Thus, users who have
been using the older calibrations can consider their wavelength
scales accurate to better than about 1 WFC pixel, but for future
applications we recommend that the updated calibrations be used. These
will be available via the ST-ECF web pages.

The wavelength solution obtained for the 1$^{st}$ and 2$^{nd}$ orders 
is a second order
polynomial in the form of: $\lambda$ = $\lambda_0$ + $\Delta\lambda_0$X$'$ +
$\Delta\lambda_1$X$'^2$, where $\lambda_0$ is the wavelength zero point in \AA,
$\Delta\lambda_0$ the  first term of the dispersion in \AA/pix and 
$\Delta\lambda_1$ the second term of the dispersion in \AA/pix$^2$. X$'$
is the distance from the X coordinate of the object in the direct image
along the spectral trace. 

For the 1$^{st}$ order, the first term of the dispersion $\Delta\lambda_0$
significantly varies across the field of view, with an amplitude of 
about 20$\%$ (of the value measured at the center of the field)
from the the top, left corner to the bottom, right
corner. Along this same direction, the wavelength zero point and 
the second term of the dispersion vary by $\sim$ 20 \AA\/ and a factor
of $\sim$ 2 respectively.

In the case of the 2$^{nd}$ order, $\Delta\lambda_0$ varies by 
about 10$\%$ of the dispersion measured at the center of the field,
along the diagonal from the top, left corner to the bottom, right
corner of the field of view. Along this same direction, the wavelength
zero point increases by $\sim$ 220 \AA.

The wavelength solutions of the positive 3$^{rd}$ and all the negative orders 
are a first order polynomial in the form of: $\lambda$ = $\lambda_0$ + 
$\Delta\lambda_0$X$'$. They exhibit the same field dependence as for the
positive 1$^{st}$ and 2$^{nd}$ orders, with $\Delta\lambda_0$ varying by
about 15$\%$ from the center to the bottom, right corner of the field of view.

It is worth noticing here that the 3$^{rd}$ orders are out of focus.
Since the focal plane is tilted with respect to the optical axis
the higher orders are more displaced from the optical axis. 
Consequently, a number of lines appear split when
compared with the template spectra of WR45 and WR96 (cf. Figure 11,
where the split lines are indicated with a vertical line).

\section{The in-orbit wavelength solutions of the HRC grism}
In the HRC grism data we detected six orders together with the 0$^{th}$
order, from the -3$^{rd}$ to the +3$^{rd}$ order. Because of the position
of WR45 across the field of view and the severe tilt of the spectra,
the $\pm$3$^{rd}$ orders are truncated and only a pair of emission lines
can be identified. We thus do not provide any wavelength solution 
for these orders.

The $\pm$1$^{st}$ and $\pm$2$^{nd}$ order spectra acquired at the
center of the field of view are plotted in Figure
12 in units of counts vs wavelength in \AA. The wavelength solutions
obtained across the HRC field of view are listed in Tables 5 and 6
(cf. Pasquali et al. 2003b).

The wavelength solution of the 1$^{st}$ order is fitted with a second
order polynomial as in the case of the WFC 1$^{st}$ order. 
The dispersion is seen varying by 8$\%$ between the top, left corner to the 
bottom, right corner of the field of view. In this case, the field dependence 
follows the same diagonal as for the WFC, only in the opposite direction. The 
wavelength zero point also varies by 0.2$\%$ between these two edges of
the field. 

A first order polynomial traces the wavelength solution of the grism
2$^{nd}$ order. The dispersion varies again
by $\sim$ 8$\%$ from the top, left corner to the
bottom, right corner of the field, and it is accompanied
by a variation of 0.3$\%$ in the wavelength zero point.  

The -1$^{st}$ and -2$^{nd}$ orders are detected only in two positions, at
the center and the bottom, right corner of the field.

\section{The grism flux calibration}
The flux calibration of the ACS grism requires the correction for
flat-field and for the total grism throughput, which takes into
account the intrinsic grism response, the quantum efficiency of the
CCD detectors and the throughput of the telescope and instrument optics.

In the case of slitless spectroscopy, the flat-field is field and wavelength
dependent at the same time. Indeed, depending on the coordinates of a source
in the direct image, the same pixel in the grism image is exposed to different 
wavelengths during different telescope pointings and grism acquisitions.
A grism flat-field can be then constructed via interpolation of the 
flat-fields available for different imaging filters at any pixel position 
and wavelength, i.e. what is refereed
to as a flat-field cube. For this purpose, we used a 3$^{rd}$ order polynomial
as a function of pixel coordinates and wavelength, which is directly used by
the extraction package aXe (Pirzkal et al. 2003a,b).
Technically, the correction for flat-field is applied following the
wavelength calibration of a spectrum: only at this point,
it is possible to extract
a monodimensional flat-field from the flat-field cube in correspondence with the 
source position in the direct image and the wavelength range of the source
spectrum.  

Ideally, the grism flat-field cube should be constructed from the flat-fields of
narrow-band filters, in order to better sample the wavelength space of the
grism spectra and therefore to perform an accurate flux calibration.
Practically, the available grism flat-field cubes (for the WFC and the HRC at 
http://www.stecf.org/instrument/acs)
are based on in-orbit flat-fields for broad-band filters which describe the
large scale variations in the CCD response up to $\lambda \sim$ 9000 \AA. 

In the case of the WFC, taking the broad-band filter flat-fields at face 
value generates a grism 
flat-field cube which produces flux discrepancies among the grism spectra of
the standard star GD153 (the same occurs for G191B2B) observed at different
positions across the field of view. These differences in flux
possibly arise from a large-scale flat properties of the grism images
that are different from those seen in direct images. In addition, since 
the imaging
flat-fields are designed to give a flat image of the sky,
their application to a
grism image introduces a correction for the geometric distortions 
which is already been taking into account in the wavelength calibration
of the extracted spectra. Consequently, the differences in flux among
the standard star spectra taken in different positions were  fitted with a 2D
surface which was used to generate a {\it corrected} grism flat-field cube,
able to produce spectra of the same star which agree within 1$\%$ between 
6000 \AA\/
and 9000 \AA\/ and 2 - 3$\%$ at redder wavelengths, all across the field 
of view of the WFC (Pirzkal et al. 2003a,b, Walsh \& Pirzkal 2005).

In the case of the HRC, the standard star GD153 was observed only in three
positions out of the five which were adopted for the wavelength calibration.
A simple grism flat-field cube, without any correction as applied to the WFC
cube, generates spectra whose fluxes already agree within 2 $\%$, most 
likely because of the much lower degree of geometric distortions in the HRC
with respect to the WFC (Pirzkal et al. 2003b).  

The spectra extracted for GD153 and G191B2B, calibrated in wavelength and
properly corrected for flat-field and CCD gain, were normalized by the
exposure time, 
and divided by the STSDAS template spectrum
calibrated in flux. The resulting sensitivity functions (in e$^-$ pix$^{-1}$
\AA\/ per erg cm$^{-2}$ \AA$^{-1}$) were averaged  into a single grism response
curve independent of the position across the chip and the standard deviation
of the mean was assumed to be the error on the grism throughput. The 
errors in the absolute flux calibration are then 3$\%$ between 5000 \AA\/ and 
9000 \AA\/ and 20$\%$ past 10000 \AA. The mean
WFC/HRC grism responses obtained for the 1$^{st}$ order are plotted in 
Figure 13. 
They are delivered with the aXe package and can be found at 
http://www.stecf.org/software/aXe together with the grism higher order
responses.

A secondary effect in the WFC/HRC grism spectra is fringing, due to 
the interference between the incident light and the light reflected at
the interfaces between the thin layers of the CCDs. The modeling of
fringes is possible once the layer composition of the CCD detector, the
thickness and material of these layers are known. In the case of ACS,
these parameters
are not available (the WFC CCDs have proprietary construction), and 
are estimated by comparison of a
fringe model with narrow-band filter flat-fields.

Walsh et al. (2003) have found that the peak-to-peak amplitude of the fringes 
in the HRC ($\simeq$ 27$\%$) is reproduced by assuming a CCD model with 
four layers: the Si top layer 12.5 - 16.0 $\mu$m thick,
the second SiO$_2$ layer
0.26 $\mu$m in thickness, the third Si$_3$N$_4$ and the fourth Si layers
with a 0.23 $\mu$m and 0.83 $\mu$m thickness respectively. Similarly,
the WFC fringes are characterized by a peak-to-peak amplitude of $\simeq$ 24$\%$ 
which is reproduced by the same four layers CCD model adopted for the HRC,
only with different thicknesses: 12.6 - 17.0 $\mu$m for the top layer,
1.14 $\mu$m, 0.90 $\mu$m and 3.08 $\mu$m for the second, third and fourth
layer respectively. 

The predictions of the above CCD models can be use to correct the narrow-band
filter flat-fields for fringing: the net result is a decrease of the
fringes peak-to-peak amplitude of about four.

Does fringing affect grism spectra? The incident light on the grism is
convolved with the instrument PSF along both the dispersion and cross-dispersed 
axes. In the case of the WFC, the FWHM of the
PSF is $\simeq$ 2.5 pixels, or 100 \AA\/ assuming $\Delta\lambda_0$ =
40 \AA. The fringes period is $\sim$70 \AA, therefore the PSF has the net
effect of smoothing out the fringes envelope, by a factor of six as 
SLIM simulations show. Therefore, fringing in the WFC grism spectra is
practically negligible. Since the grism dispersion is higher when coupled
with the HRC ($\Delta\lambda_0$ $\simeq$ 24 \AA) and the fringes period 
is 65 \AA, some residual fringing may still be observed in HRC grism spectra 
(Walsh et al. 2003).

At the present time, a routine for the fringing correction is being
planned for aXe.

\section{From the calibrations to the users}

The in-orbit calibrations of the ACS grism served the purpose to characterize
its physical properties (i.e. spectral tilt) and the wavelength solution
for each of its detected orders. Equally important, the calibrations were 
performed in selected positions across the WFC and HRC fields of view to
verify the field dependence of the grism properties. The next step is therefore
to be able to extract and calibrate a spectrum in wavelength and flux 
{\it at any position in the field of view}. This is achieved by interpolating
each of the quantities derived in the previous sections across the chip as a
function of the (X,Y) coordinates of the calibration stars, WR45 and WR96,
in the direct image. Each 
interpolation describes the smooth spatial variation of the grism properties.

Specifically, the quantities that need to be interpolated are: the spectral
tilt, the Y-offset and, for each order, the dispersion (the second term of
the dispersion when available) and wavelength zero point. In the case of the
WFC, the observed positions allowed us to parameterize the grism field dependence
with second-order 2D polymonials, which, in the case of the grism 1$^{st}$ order,
are characterized by an error of 
0$^o$.03 in the spectral tilt, 0.10 pixels in the Y-offset, 10 and 5 \AA\/ in
the wavelength zero point in Chip 1 and 2, respectively, 0.6$\%$ and 0.4$\%$ in the
first ($\Delta\lambda_0$) term of the dispersion
in Chip 1 and 2, respectively. The second ($\Delta\lambda_1$) term is characterized
by a r.m.s of 27$\%$ and 22$\%$ in Chip 1 and 2, respectively (Larsen \& Walsh 2005).
These interpolations were used to extract the spectra of the Planetary Nebula SMP-81, which 
in turn confirmed the consistency of the grism wavelength calibration. 

Since only 5 positions were acquired across the HRC field of view, we 
interpolated the grism properties with a first-order 2D polynomial, 
which, for the 1$^{st}$ order, has errors 
of 0$^o$.0005 in the spectral tilt, 0.4 pixels in the
Y-offset, 4.8 \AA\/ (i.e. 0.2 pixels) in the wavelength zero point and 
0.2$\%$ and 10$\%$ in the first ($\Delta\lambda_0$) and second ($\Delta\lambda_1$) 
terms of the dispersion (Pasquali et al. 2003b).  

The coefficients of the 2D polynomials fitting the field dependence of the grism
properties are stored in {\it configuration files} (one for each WFC chip and one
for HRC)
which are used by aXe and can be retrieved at
http://www.stecf.org/software/aXe. The configuration files contain three other
fundamental keywords: the name of the flat-field cube to be applied for the 
flat-field correction, the name of the sensitivity function for the flux 
calibration and the length of the orders in pixels.

To test the consistency of our interpolation, we re-extracted the spectra
of WR45 and WR96 with aXe and the configuration files described above. 
Figure 14 shows the results for the different pointings across Chip 1
of the WFC. The spectra coincide within 0.4 [$\sim$ 16 \AA, or 0.2 pixels 
($\sim$ 5 \AA) in the case of the HRC].

\section{Concluding remarks}

Observations with the ACS grism require science targets to be imaged
through a filter (direct image) and the grism (grism image). The 
reason is that the target position in the direct image defines
the zero point in the wavelength calibration and the flat-field
correction for the flux calibration of a spectrum in the
grism image.

Calibration data have shown that the object position in the direct 
image is independent of the filter used for imaging and is stable
within 25 mas (0.5 pixels) when multiple acquisitions are performed within the 
same orbit. Direct images taken of the same target with the same
POSTARG values during different orbits are instead stable within
100 mas (2 pixels). While a shift of $<$ 0.5 pixels (i.e.
about 20 \AA\/ in the grism first order spectrum) 
is within the uncertainties of
the wavelength calibration, an offset of 2 pixels introduces a
systematic error of $\sim$ 80 \AA\/ 
in the wavelength calibration of the grism 1$^{st}$
order. It is thus advisable to always acquire a direct image 
of the targeted field together with its grism images.
In the case of multiple exposures of the same field with
dither shifts, it suffices to have one direct image corresponding
to one of the grism images and to use the 0$^{th}$ orders to confirm 
the telescope shifts (e.g. this technique was applied by Pirzkal
et al. 2004 to the GRAPES data).

Direct images can be combined to increase the S/N threshold for the detection of
faint objects, but the positions of the objects measured in the 
final drizzled version have to be related to positions in the
raw, geometrically distorted frame for a correct use of aXe
(cf. Pirzkal et al. 2004). A spectroscopic drizzle task has been
implemented in the latest version of aXe (aXe-1.4, K\"ummel et al.
2004), which extracts 2D stamp images of a source spectrum appearing
in multiple grism images, fully calibrated in wavelength and corrected
for flat field. In the most general case, the source has been imaged
in different positions across the field of view, therefore, because
of the field dependence of the spectral dispersion, its spectra are
characterized by different spectral dispersions. The aXe drizzle
corrects the tilt of the individual spectral stamp image, so that
its dispersion and the cross-dispersed axes are parallel
to the x and y axes, respectively. It also makes the wavelength
scale and the pixel scale in the cross-dispersed direction the same in
all the spectral stamp images of the source. These drizzled stamp
images are then coadded, and the final 1D spectrum of the source is
extracted and flux calibrated, using a weighting scheme based on the
S/N variations as provided by the drizzle weights (hence proportional
to the different exposure times of the observations). Users can set
the wavelength scale and the pixel scale in the cross-dispersed direction
in the aXe configuration file; this option allows to enhance the
resolution (spectral and angular) of the extracted spectra if the original 
data were taken with subpixel stepping.

The ACS slitless grism is equivalent to a ground-based multi-object 
spectrograph. The absence of a slit, though,
affects the grism resolution, which the object size and inclination with 
respect to the dispersion axis can easily decrease (cf. Sect. 3). This effect 
can be modeled by, for
example, deconvolving the spectrum by the spatial profile of the source,
once the latter is known as a function of wavelength.
In addition, the large extension of the grism orders may cause
spectra of two adjacent objects to overlap. In its present form aXe
provides a quantitative estimate of contamination but it 
does not correct for spectra contamination. Therefore it is suggested
to acquire data using different telescope roll angles so to vary
the direction of the dispersion axis and reduce the amount of spectra overlap.
Also, orienting the telescope so that the dispersion axis is aligned with
the object minor axis, certainly maximizes the spectral resolution 
achievable for that target. 
A final note has to be spent on the background level. Since each pixel
of the spectrum sees the background emission integrated over the whole
spectral range of the grism, the use of a high S/N background image,
properly scaled to the exposure time of the data, will improve the
S/N in the extracted spectra of faint sources.

As already mentioned before, ACS slitless grism spectroscopy has 
extensively been carried out by large programs such as APPLES 
(ACS Pure Parallel Ly$\alpha$ Emission Survey, PI Rhoads, ID 9482),
GRAPES (Grism ACS Program for Extragalactic Science, PI Malhotra,
ID 9793) and PEARS (Probing Evolution and Reionization 
Spectroscopically, PI Malhotra, ID 10530). GRAPES in particular has
proved that the ACS grism can detect the continuum emission of sources
as faint as z$_{AB}$ = 27.2 mag and up to $z \sim$ 7 using 40 orbits
(Pirzkal et al. 2004). This is the deepest slitless spectroscopy survey 
to date, even when compared with ground-based telescopes. It has allowed
to characterize several types of galaxy population: Lyman break and Ly$\alpha$ 
galaxies at 4 $< z <$ 7 (Rhoads et al. 2005), old stellar populations at 0.5 $< z <$ 2.5 
(Daddi et al. 2005, Pasquali et al. 2005), and emission line galaxies at $z <$ 1.5
(Pirzkal et al. 2005b), often at a luminosity level fainter than previously reached.
The success of these surveys clearly indicate that the ACS grism spectroscopy 
provides important insights on galaxy evolution at high redshift.

\acknowledgments
We would like to thank F. Paresce and
F. van den Bosch for inspiring discussions and comments, and an anonymous
referee for improving the paper. We acknowledge
support from F. Pierfederici and M. Dolenski in querying the ST-ECF
Archive, and help from S. Baggett and P. Goudfrooj on HST
documentation.

\clearpage






\begin{figure*}
\epsscale{1.0}
\plotone{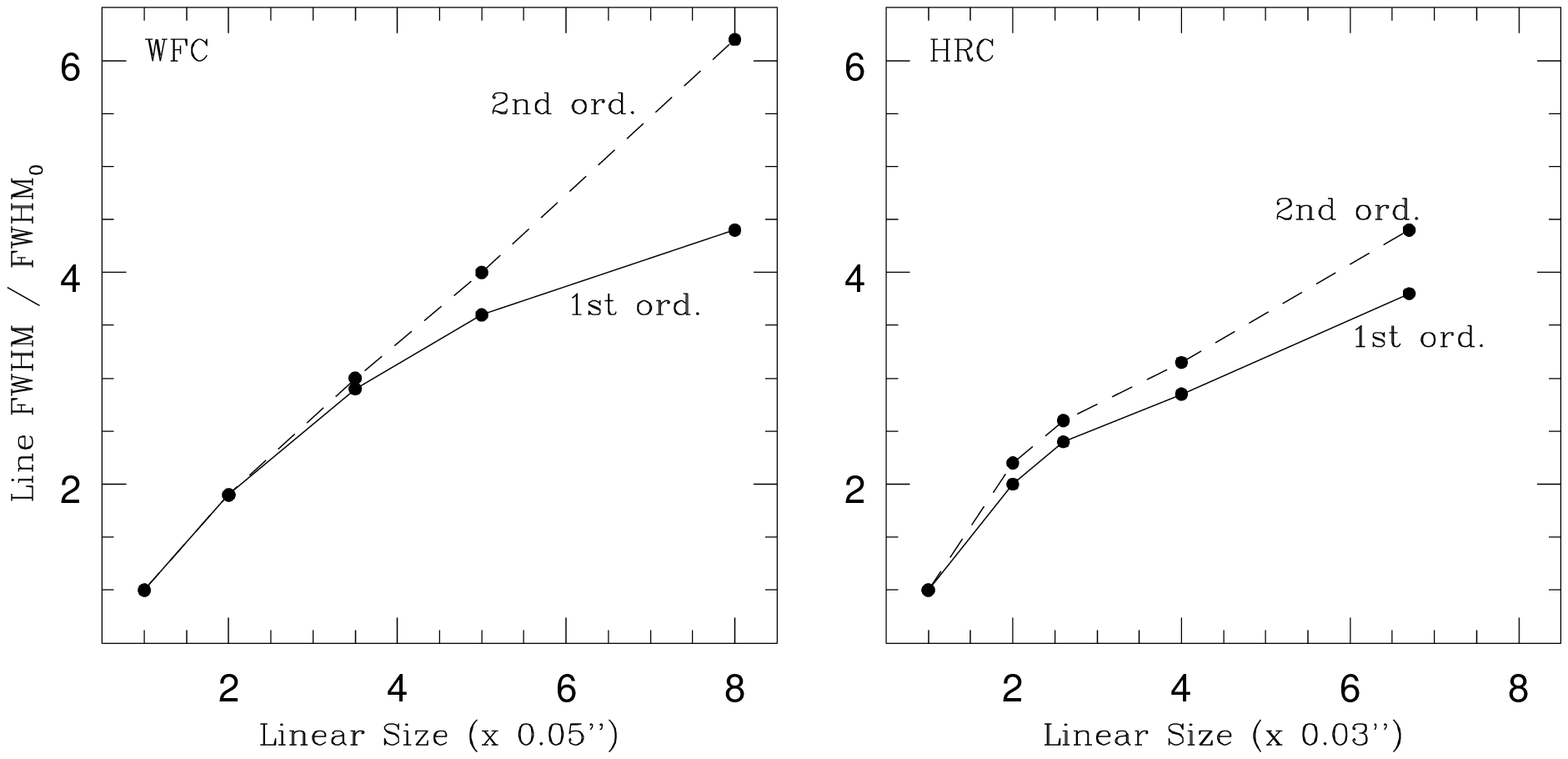}
\caption{The spectral resolution of the WFC and HRC grism as a function
of the object size. The solid line represents 1$^{st}$ order spectra and the
dashed line corresponds to 2$^{nd}$ order spectra. The object diameter 
along the dispersion axis is
plotted in abscissa as a multiple of the PSF FWHM in arcsecs. The line
FWHM is represented in the ordinate axis as a multiple of the mean FWHM
in \AA\ computed from the lines in the spectrum of a source 0$''$.05
in diameter.}
\end{figure*}


\begin{figure*}
\epsscale{1.0}
\plotone{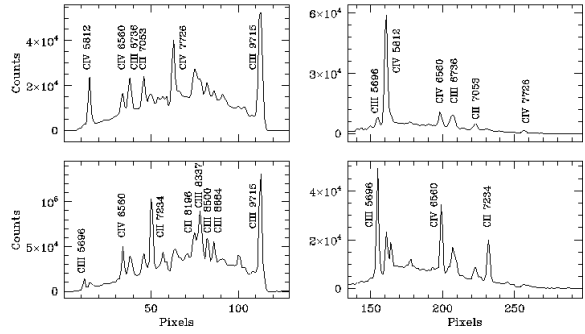}
\caption{The SLIM WFC spectra for WR stars of spectral template WC6 (top)
and WC8 (bottom). First  orders are plotted
in the left column and refer to an exposure time of 10 s. The second order
spectra are shown in the right column and have been computed for an integration
time of 60 s.}
\end{figure*}

\begin{figure*}
\epsscale{1.0}
\plotone{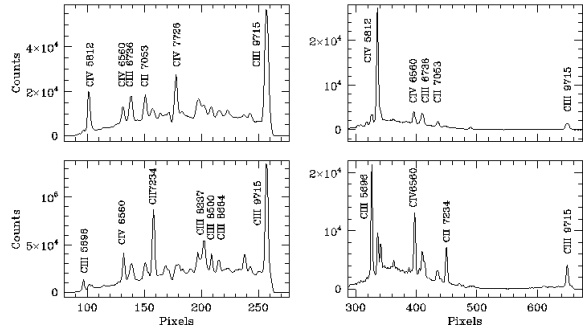}
\caption{As in Figure 2 but for the HRC.
First  orders are plotted
in the left column and refer to an exposure time of 20 s. The second order
spectra are shown in the right column and have been computed for an integration
time of 60 s.}
\end{figure*}

\begin{figure*}
\epsscale{1.0}
\plotone{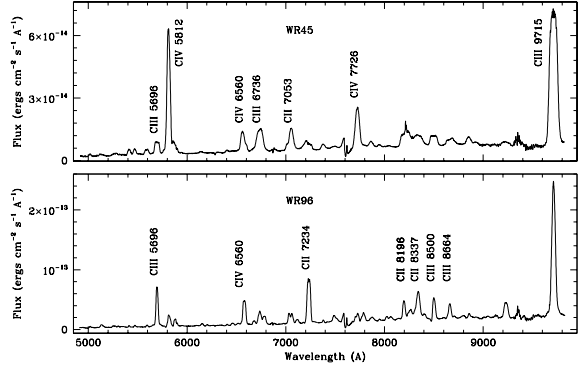}
\caption{The NTT/EMMI spectra of WR45 (top) and WR96 (bottom).}
\end{figure*}

\begin{figure*}
\plotone{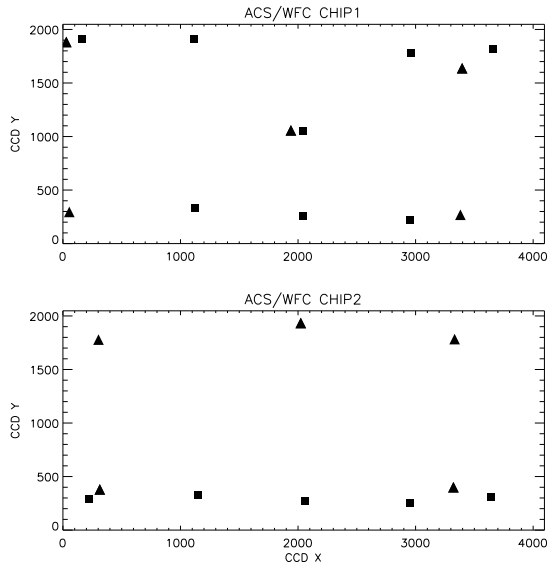}
\caption{Observations of WR96. The object positions in the direct image 
used for programs 9568 (INTERIM) and 10058 (Cycle 12) are shown with 
triangles and boxes, respectively.}
\label{fig:pos}
\end{figure*}

\begin{figure*}
\epsscale{0.9}
\plotone{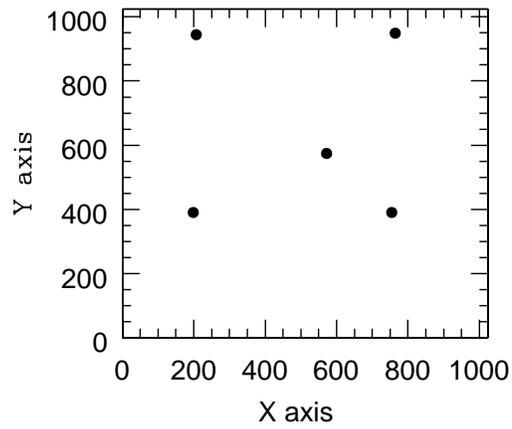}
\caption{The map of the pointings (i.e. object positions in the direct image) 
used for WR45 and WR96 across the field of view
of the HRC.}
\end{figure*}

\begin{figure*}
\epsscale{1.0}
\plotone{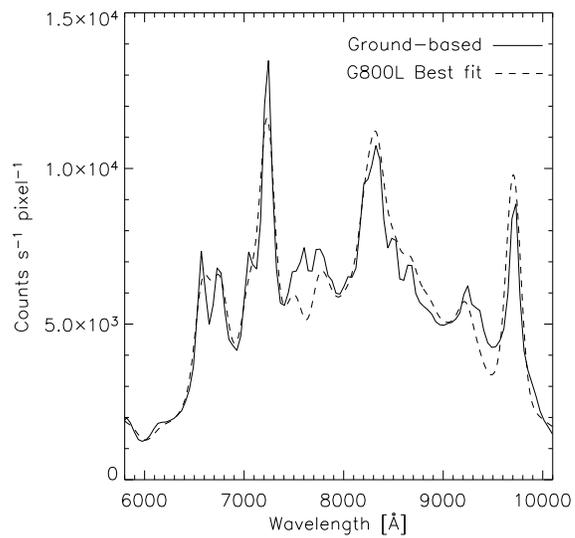}
\caption{The smoothed NTT spectrum of WR96 and the ACS/WFC grism spectrum
  for the best fitting wavelength solution.}
\label{fig:fitfig}
\end{figure*}

\begin{figure*}
\epsscale{1.0}
\plotone{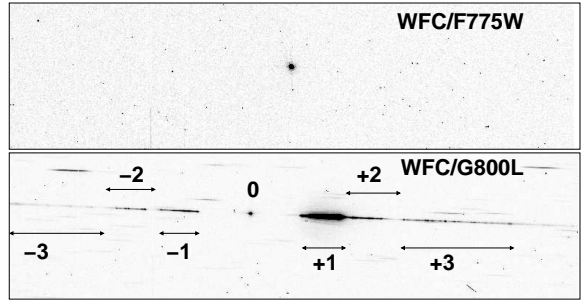}
\caption{The 2D grism spectrum of WR96 taken with the WFC (bottom panel), where
seven orders, from the negative to the positive third, have been identified.
The upper panel shows the position of the object in the direct image at
which the grism spectrum was acquired. The size of both images
is 1500 $\times$ 400 pixels.}
\end{figure*}

\begin{figure*}
\epsscale{0.9}
\plotone{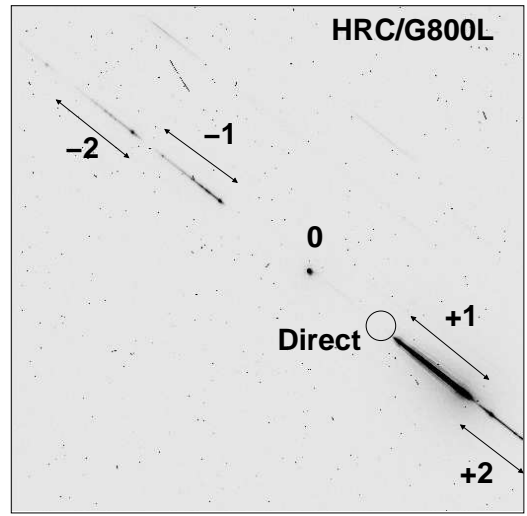}
\caption{As in Figure 8, but for the HRC. The position of
the object in the direct image is represented with a
circle. The image size is 1024 $\times$ 1024 pixels.}
\end{figure*}

\begin{figure*}
\epsscale{1.0}
\plotone{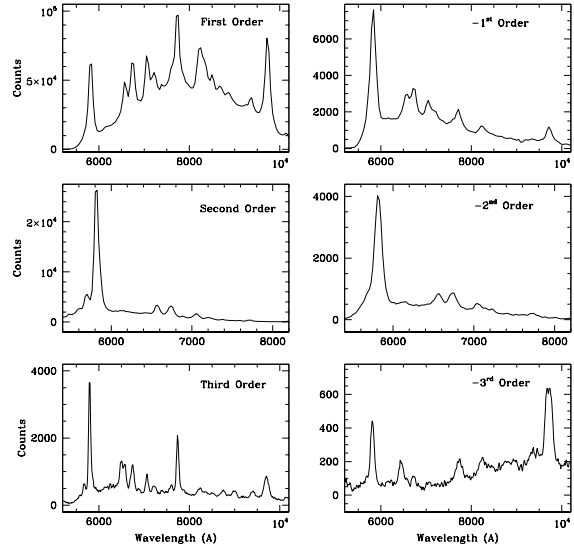}
\caption{The spectra of the six grism orders detected in the WFC data
and obtained for WR45.}
\end{figure*}

\begin{figure*}
\epsscale{1.0}
\plotone{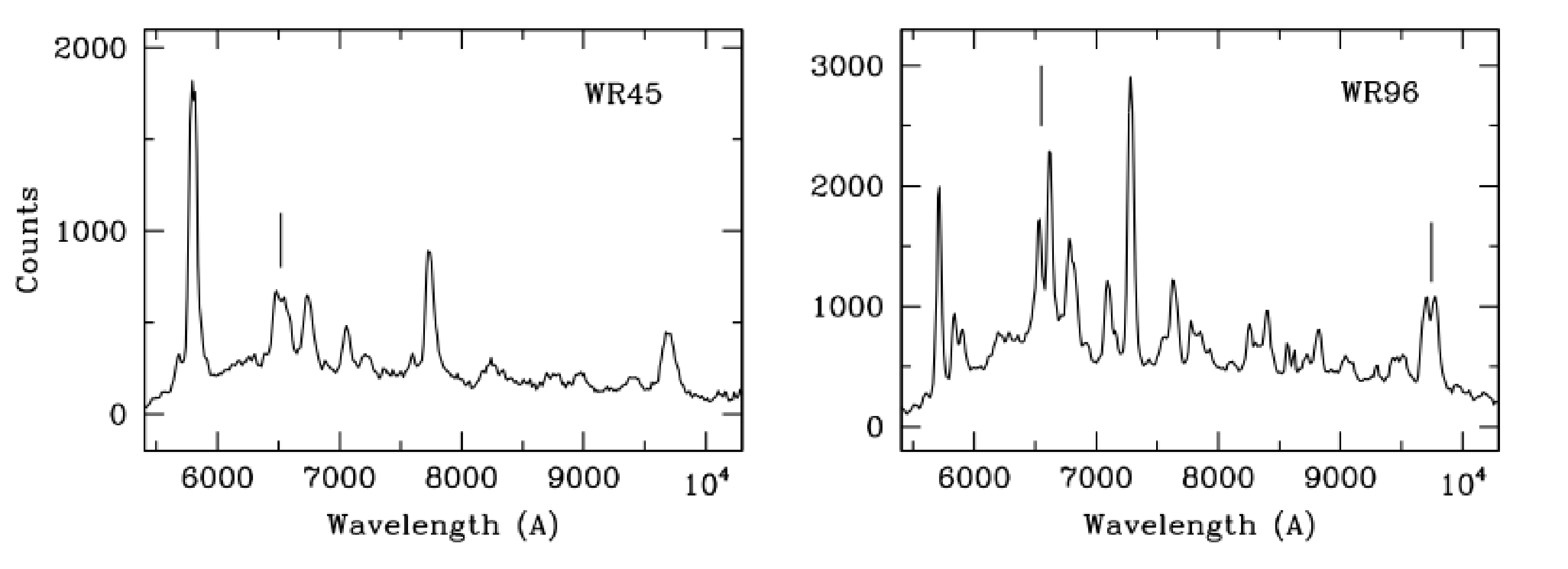}
\caption{The third order spectra of WR45 and WR96. The split lines are indicated
with a vertical line.}
\end{figure*}

\begin{figure*}
\epsscale{1.0}
\plotone{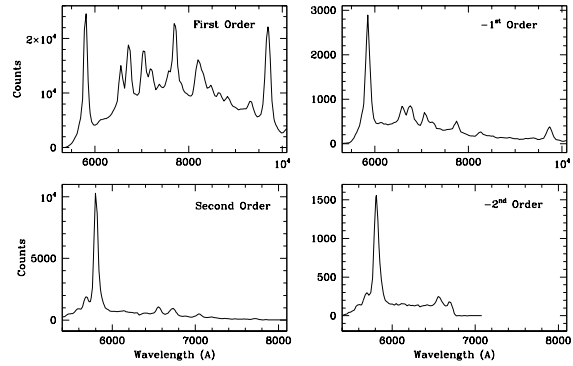}
\caption{The WR45 spectra of the four orders detected and calibrated in the 
HRC data.}
\end{figure*}

\begin{figure*}
\epsscale{1.0}
\plotone{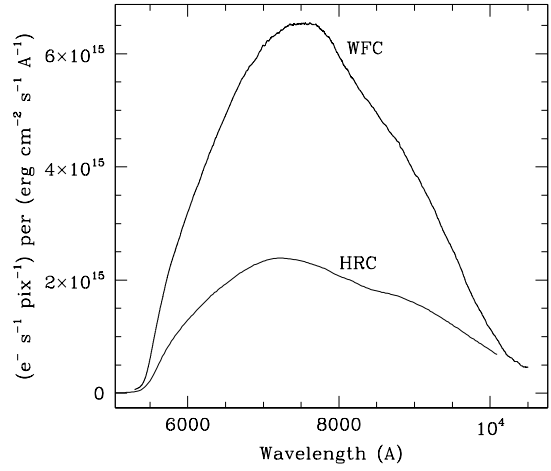}
\caption{The sensitivity functions of the grism 1$^{st}$ order
when coupled with the WFC and the HRC.}
\end{figure*}

\begin{figure*}
\epsscale{1.0}
\plotone{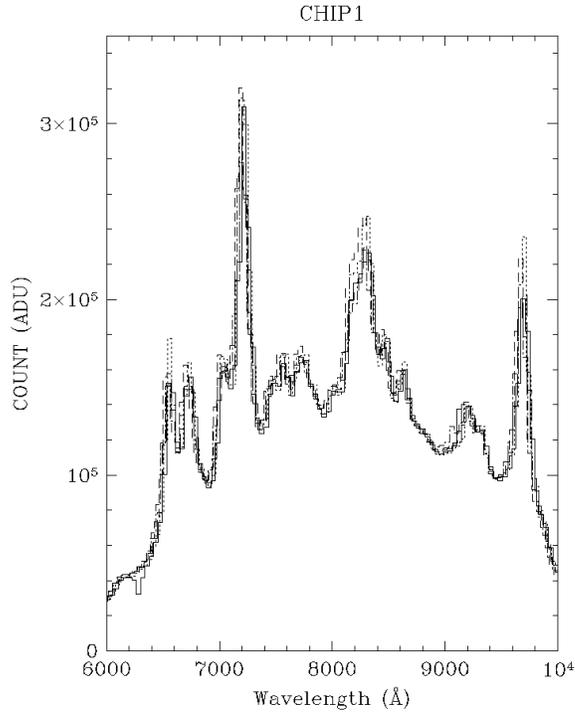}
\caption{The spectra of WR96 extracted in  5 positions across 
Chip 1 of the Wide Field Channel.}
\end{figure*}








\clearpage

\begin{deluxetable}{cccccc}
\tablecaption{Basic parameters of the selected WR stars, WR45 and WR96.}
\tablewidth{0pt}
\tablehead{
\colhead{Star} & \colhead{Spectral type} & \colhead{RA(2000)} & 
\colhead{DEC(2000)} & \colhead{V} & \colhead{$v_{\infty}$ (km s$^{-1}$)}
}
\startdata
WR45 & WC6 & 11:38:05.2 & -62:16:01 & 14.80 & 2100 \\
WR96 & WC9 & 17:36:24.2 & -32:54:29 & 14.14 & 1100 \\
\enddata
\end{deluxetable}

\begin{deluxetable}{lrrr}
\tablecaption{The exposure times adopted for each star
and spectral element.} 
\tablewidth{0pt}
\tablehead{
\colhead{Target} & \colhead{F625W} & \colhead{F775W} &
\colhead{G800L}
}
\startdata
WFC:   &      &     &    \\
WR45   & 1 s  & 1 s & 20 s\\
WR96   &      & 1 s & 20 s\\
WR96   &      & 1 s & 15 s\\
LMC-SMP-81 &  & 15 s & 200 s\\
G191B2B&      & 1 s & 15 s\\
GD153  &      & 2 s & 60 s\\
       &      &     & \\
HRC:   &      &     &\\
WR45   &      & 3 s & 60 s\\
GD153  &      & 6 s & 180 s\\
\enddata
\end{deluxetable}

\begin{deluxetable}{ccccccccc}
\rotate
\tabletypesize{\scriptsize}
\tablecaption{\label{tab:wr96pos}The wavelength solutions of 
the 1$^{st}$, 2$^{nd}$ and 3$^{rd}$ 
orders, determined for the different pointings across the field of view 
of the WFC.}
\tablewidth{0pt}
\tablehead{
\colhead{Position} & \colhead{First order} & \colhead{} & \colhead{} &
\colhead{Second order} & \colhead{} & \colhead{} & \colhead{Third order} & \colhead{} \\
\colhead{} & \colhead{$\lambda_0$} & \colhead{$\Delta\lambda_0$} & \colhead{$\Delta\lambda_1$}
           & \colhead{$\lambda_0$} & \colhead{$\Delta\lambda_0$} & \colhead{$\Delta\lambda_1$}
           & \colhead{$\lambda_0$} & \colhead{$\Delta\lambda_0$}
}
\startdata
WFC/CHIP1     & & & & & & & & \\
(29, 1883)   & 4762.17 & 35.50 &  0.00414 & 2546.53 & 16.91 & 0.00306 & 1529.99 & 12.46 \\
(29, 1883)   & 4763.12 & 35.49 &  0.00432 & 2550.02 & 16.88 & 0.00314 & 1537.05 & 12.44 \\
(54, 296)    & 4769.58 & 37.27 &  0.00480 & 2634.18 & 16.91 & 0.00549 & 1537.22 & 13.06 \\
(1939, 1058) & 4733.32 & 39.93 & -0.00007 & 2628.73 & 17.82 & 0.00556 & 1541.92 & 13.64 \\
(1940, 1057) & 4738.62 & 39.98 & -0.00025 & 2636.27 & 17.76 & 0.00570 & 1540.71 & 13.65 \\
(3396, 1638) & 4765.84 & 40.21 &  0.00542 & 2652.97 & 18.11 & 0.00657 & 1552.87 & 14.03 \\
(3396, 1638) & 4764.46 & 40.24 &  0.00519 & 2653.43 & 18.08 & 0.00666 & 1538.95 & 14.07 \\
(3381, 269)  & 4776.26 & 41.87 &  0.00771 & 2633.99 & 19.22 & 0.00644 & 1543.18 & 14.71 \\
(2950, 223)  & 4778.30 & 41.81 &  0.00414 & 2708.75 & 18.15 & 0.00865 & 1546.97 & 14.53 \\
(3662, 1820) & 4751.05 & 40.61 &  0.00373 & 2577.85 & 18.99 & 0.00437 &   -     &   -   \\
(2959, 1778) & 4769.01 & 39.67 &  0.00367 & 2629.52 & 18.08 & 0.00552 & 1565.00 & 13.75 \\
(163, 1910)  & 4756.89 & 35.98 &  0.00279 & 2508.35 & 17.42 & 0.00202 & 1564.33 & 12.44 \\
(1111, 1915) & 4768.77 & 36.99 &  0.00420 & 2607.01 & 17.06 & 0.00446 & 1575.82 & 12.83 \\
(1125,  337) & 4776.17 & 38.97 &  0.00352 & 2657.27 & 17.41 & 0.00635 & 1553.18 & 13.55 \\
(2044,  257) & 4769.19 & 40.56 &  0.00302 & 2710.59 & 17.52 & 0.00817 & 1547.22 & 14.05 \\
(2045, 1054) & 4774.98 & 39.24 &  0.00451 & 2665.68 & 17.50 & 0.00664 & 1559.31 & 13.66 \\
WFC/CHIP2     & & & & & & & & \\
(3332, 1784) & 4746.52 & 43.32 &  0.00334 & 2699.07 & 18.72 & 0.00917 & 1548.27 & 14.94 \\
(3322,  400) & 4777.15 & 44.73 &  0.00806 & 2728.65 & 19.34 & 0.01084 & 1536.41 & 15.71 \\
(3322,  400) & 4782.92 & 44.67 &  0.00852 & 2727.76 & 19.36 & 0.01081 & 1544.99 & 15.69 \\
(303, 1779)  & 4761.05 & 38.56 &  0.00327 & 2609.03 & 17.63 & 0.00511 & 1389.09 & 13.77 \\
(314,  380)  & 4782.26 & 39.77 &  0.00721 & 2614.51 & 18.40 & 0.00555 & 1532.86 & 14.04 \\
(314,  380)  & 4772.47 & 39.86 &  0.00669 & 2665.29 & 17.79 & 0.00720 & 1553.76 & 13.98 \\
(2023, 1934) & 4774.60 & 40.78 &  0.00462 & 2701.40 & 17.79 & 0.00816 & 1551.22 & 14.20 \\
(2024, 1934) & 4775.98 & 40.94 &  0.00348 & 2698.68 & 17.85 & 0.00805 & 1563.61 & 14.18 \\
(3642,  309) & 4758.08 & 45.90 &  0.00491 & 2730.46 & 19.61 & 0.01124 & 1541.64 & 15.93 \\
(2954,  253) & 4774.10 & 44.81 &  0.00476 & 2764.58 & 18.74 & 0.01207 & 1555.78 & 15.54 \\
(225,  289)  & 4775.20 & 39.98 &  0.00561 & 2607.48 & 18.45 & 0.00548 & 1529.12 & 14.04 \\
(1151,  327) & 4786.96 & 41.63 &  0.00497 & 2646.24 & 18.83 & 0.00664 & 1534.97 & 14.55 \\
(2059,  268) & 4788.72 & 43.10 &  0.00545 & 2672.69 & 19.25 & 0.00793 & 1559.76 & 15.02 \\
\enddata
\end{deluxetable}

\begin{deluxetable}{ccccccc}
\rotate
\tabletypesize{\scriptsize}
\tablecaption{\label{tab:wr96neg}The wavelength solutions of the 
-1$^{st}$, -2$^{nd}$ and -3$^{rd}$ 
orders, determined for the different pointings across the field of view 
of the WFC.}
\tablewidth{0pt}
\tablehead{
\colhead{Position} & \colhead{Negative First order} & \colhead{} &
\colhead{Negative Second order} & \colhead{} & \colhead{Negative Third order} & \colhead{} \\
\colhead{} & \colhead{$\lambda_0$} & \colhead{$\Delta\lambda_0$} 
           & \colhead{$\lambda_0$} & \colhead{$\Delta\lambda_0$} 
           & \colhead{$\lambda_0$} & \colhead{$\Delta\lambda_0$}
}
\startdata
WFC/CHIP1    &          &        &          &        &          & \\
(29, 1883)   &    -     &    -   &   -      &    -   &    -     &    -   \\
(29, 1883)   &    -     &    -   &   -      &    -   &    -     &    -   \\
(54, 296)    &    -     &    -   &   -      &    -   &    -     &    -   \\
(1939, 1058) & -4965.81 & -40.61 & -2317.55 & -19.66 & -1472.93 & -12.91 \\
(1940, 1057) & -5045.09 & -40.83 & -2319.40 & -19.65 & -1472.69 & -12.90 \\
(3396, 1638) & -5069.28 & -42.43 & -2355.43 & -20.47 & -1531.79 & -13.49 \\
(3396, 1638) & -5081.47 & -42.47 & -2374.06 & -20.52 & -1565.35 & -13.54 \\
(3381, 269)  & -5123.16 & -44.59 & -2340.16 & -21.39 & -1530.69 & -14.13 \\
(2950, 223)  & -5115.70 & -43.85 & -2376.59 & -21.14 & -1485.96 & -13.84 \\
(3662, 1820) & -5094.49 & -42.70 & -2377.55 & -20.61 & -1490.10 & -13.50 \\
(2959, 1778) & -5077.32 & -41.56 & -2359.10 & -20.06 & -1554.45 & -13.26 \\
(163, 1910)  &    -     &   -    &   -      &    -   &    -     &    -   \\
(1111, 1915) & -5046.21 & -38.53 & -2334.57 & -18.56 & -1287.35 & -11.90 \\
(1125,  337) & -5048.17 & -40.47 & -2348.77 & -19.53 & -1565.10 & -12.91 \\
(2044,  257) & -5110.14 & -42.33 & -2367.31 & -20.38 & -1494.76 & -13.36 \\
(2045, 1054) & -5091.67 & -41.17 & -2359.96 & -19.84 & -1494.91 & -13.00 \\
WFC/CHIP2    &          &        &          &        &          & \\
(3332, 1784) & -5094.14 & -45.24 & -2375.18 & -21.84 & -1623.83 & -14.51 \\
(3322,  400) & -5129.59 & -47.57 & -2377.74 & -22.90 & -1666.43 & -15.31 \\
(3322,  400) & -5096.84 & -47.42 & -2381.09 & -22.91 & -1621.90 & -15.23 \\
(303, 1779)  &    -     &   -    &   -      &    -   &    -     &    -   \\
(314,  380)  &    -     &   -    &   -      &    -   &    -     &    -   \\
(314,  380)  &    -     &   -    &   -      &    -   &    -     &    -   \\
(2023, 1934) & -5059.91 & -42.64 & -2352.79 & -20.57 & -1460.89 & -13.46 \\
(2024, 1934) & -5069.51 & -42.64 & -2356.22 & -20.57 & -1478.50 & -13.48 \\
(3642,  309) & -5310.33 & -49.11 & -2381.69 & -23.31 & -1482.93 & -15.25 \\
(2954,  253) & -5111.24 & -47.01 & -2548.88 & -23.09 & -1456.04 & -14.79 \\
(225,  289)  &    -     &   -    &   -      &    -   &    -     &    -   \\
(1151,  327) & -5056.06 & -43.33 & -2352.71 & -20.91 & -1522.75 & -13.77 \\
(2059,  268) & -5109.74 & -45.28 & -2369.10 & -21.81 & -1463.58 & -14.24 \\
\enddata
\end{deluxetable}

\begin{deluxetable}{cccccc}
\tablecaption{The wavelength solutions of the 1$^{st}$, 2$^{nd}$ and 3$^{rd}$ orders, determined
for the different pointings across the field of view of the HRC.}
\tablewidth{0pt}
\tablehead{
\colhead{Position} & \colhead{First order} & \colhead{} & \colhead{} &
\colhead{Second order} & \colhead{}\\
\colhead{} & \colhead{$\lambda_0$} & \colhead{$\Delta\lambda_0$} &
\colhead{$\Delta\lambda_1$} & \colhead{$\lambda_0$} & \colhead{$\Delta\lambda_0$} 
}
\startdata
(572, 574) & 4755.91  & 23.86  & 0.0018  & 2475.29  & 12.00 \\ 
(207, 944) & 4749.06  & 24.59  & 0.0019  & 2471.58  & 12.35 \\
(754, 391) & 4759.68  & 23.51  & 0.0019  & 2480.18  & 11.82 \\
(199, 391) & 4756.92  & 23.87  & 0.0018  & 2476.84  & 11.99 \\
(764, 949) & 4753.80  & 24.19  & 0.0021  & 2471.41  & 12.19 \\
\enddata
\end{deluxetable}

\begin{deluxetable}{ccccc}
\tablecaption{The wavelength solutions of the -1$^{st}$, -2$^{nd}$ and -3$^{rd}$ orders, determined
for the different pointings across the field of view of the HRC.}
\tablewidth{0pt}
\tablehead{
\colhead{Position} & \colhead{Negative first order} & \colhead{} &
\colhead{Negative second order} & \colhead{} \\
\colhead{} & \colhead{$\lambda_0$} & \colhead{$\Delta\lambda_0$} &
\colhead{$\lambda_0$} & \colhead{$\Delta\lambda_0$}
}
\startdata
(572, 574) & -5182.09  & -26.75  & -2762.80  & -13.54  \\
(207, 944) &    -      &   -     &    -      &   -     \\
(754, 391) & -5191.30  & -26.40  & -2732.50  & -13.31  \\
(199, 391) &    -      &   -     &   -       &   -\\
(764, 949) &    -      &   -     &   -       &   -\\
\enddata
\end{deluxetable}

\end{document}